\documentclass[aps,prb,twocolumn,showpacs]{revtex4-1}

\usepackage{amsmath}
\RequirePackage{amsfonts}
\RequirePackage{amssymb}
\usepackage{color}
\usepackage[normalem]{ulem}
\RequirePackage[british]{babel}
\RequirePackage{graphicx}
\DeclareGraphicsExtensions{.eps} %
\usepackage[usenames,dvipsnames,svgnames,table]{xcolor}
\usepackage{epstopdf}
\usepackage{dcolumn}
\usepackage{bm}

\newcommand{\bequ}{\begin{equation}}
\newcommand{\eequ}{\end{equation}}
\newcommand{\bra}{\left\langle}
\newcommand{\braq} { \right|}
\newcommand{\ket}{\right\rangle}
\newcommand{\ketq}  {\left|}
\newcommand{\beque}{\begin{eqnarray*}}
\newcommand{\eeque}{\end{eqnarray*}}
\newcommand{\bfm}{\mathbf}

\begin{document}
\setcounter{secnumdepth}{2}

\title{Theory of projections with non-orthogonal basis sets: Partitioning techniques and effective Hamiltonians}

\author{M. Soriano* and J. J. Palacios}
\affiliation{Departamento de F\'{i}sica de la Materia Condensada, Instituto de F\'isica de la Materia Condensada (IFIMAC), and Instituto Nicol\'as Cabrera (INC), Universidad Aut\'{o}noma de Madrid, Campus de Cantoblanco,
28049 Madrid, Spain}
\email{maria.soriano@uam.es}

\date{\today}

\pacs{73.63.-b,73.63.Rt}

\begin{abstract}
\noindent Here we present a detailed account of the fundamental problems one encounters in projection theory when non-orthogonal basis sets are used for representation of the operators. In particular, we re-examine the use of projection operators in connection with the calculation of projected (or reduced) Green's functions and associated physical quantities such as the local density of states (LDOS), local charge, and conductance. The unavoidable ambiguity in the evaluation of the LDOS and charge is made explicit with the help of simple examples of metallic nanocontacts while the conductance, within certain obvious limits, remains invariant against the type of projection. We also examine the procedure to obtain effective Hamiltonians from reduced Green's functions. For completeness we include a comparison with results obtained with block-orthogonal basis sets where both direct and dual spaces are used.
\end{abstract}

\maketitle

\section{Introduction}
\noindent As experiments in nanoelectronics\cite{Agrait:pr:03} or molecular electronics\cite{cuevasbook} become more sophisticated, being able to reveal physical phenomena in unprecedented detail, the need for an improvement of the theoretical description of the electronic structure of the studied systems is more and more pressing. Fortunately,   on many occasions, one can readily identify an {\em active} region, i.e., a region in space where the relevant physics takes place and which may be the only one needed of a sophisticated theoretical treatment. In electronic transport problems the active region is where the largest resistance is located. For instance, when a single molecule is connected by metallic electrodes, the main source of resistance is the molecule itself\citep{Xue:jcp:01,PhysRevB.63.121104,Palacios:prb:01,Palacios01,cuevasbook}. Another prototypical example of active region is the transition metal (TM) atom at the core of organic molecules such as Phthalocyanines\citep{phtalocyanine}. When these molecules are adsorbed on metallic surfaces, low-temperature scanning tunnelling spectroscopy usually reveals a Kondo effect associated to the TM atom\citep{Franke11,strozecka:prl:12,Jacob13}.  

The partition of the Hamiltonian of systems consisting of several parts of different nature and different physical relevance is  particularly necessary when correlations are important in some of these parts and standard density functional theory (DFT) approximations do not capture the relevant physics there. DFT implementations as those found in commonly used codes\citep{SIESTA,g09} usually make no distinction in the treatment of the different parts of a physical system (DFT+U corrections, being an exception). As regards the previous examples, an accurate  calculation of the electronic structure of the contacted molecule is imperative, in particular of the HOMO-LUMO gap which is not correctly given by standard  DFT approximations\citep{PhysRevLett.95.146402,Koentopp06,PhysRevB.77.155402,PhysRevB.80.245427,PhysRevLett.97.216405}. On the contrary, the electronic structure deep into the bulk electrodes, which are typically free-electron metals, needs little attention and even simple tight-binding models can account for it\citep{Palacios:prb:01,ANT:G,Jacob11}. Following with the second example,  in order to properly describe the Kondo effect  electronic correlations need to be taken into 
consideration at the TM atom (for instance through DFT+DMFT schemes\citep{Jacob13}) while the electronic structure of the organic ligands and substrate can be safely described by standard DFT. 

In this regard one first needs to obtain an effective description of the active part. In other words, one looks for a non-Hermitian energy-dependent ``Hamiltonian" which includes the effect of the rest of the system. Once this effective Hamiltonian has been obtained, one can try to improve the electronic description or add the necessary terms to it to account for the required physics.
However, a preliminary step, with mathematical pitfalls easy to overlook, needs to be taken first. This step consists of finding a precise mathematical definition of the active region itself. This is done, in principle, through the selection of a subset of basis elements that expand the vector subspace associated to such region. 
For practical purposes many implementations of DFT make use of non-orthogonal basis sets (typically atomic orbitals) which makes this selection problematic due to the inherent ambiguity accompanying  these subsets. In other words, the vector subspace expanded by a non-orthogonal basis subset is not orthogonal to the rest of the system from which we want it unambiguously separated. 

An important effort to address these issues has been made with the development of Wannier functions \citep{PhysRevB.61.10040,PhysRevB.77.085122,PhysRevB.56.12847, PhysRevB.69.035108, RevModPhys.84.1419, 0953-8984-22-38-385601} or localized molecular orbitals \citep{:/content/aip/journal/jcp/112/4/10.1063/1.480730, PhysRevB.88.085404}. While constructing orthogonal basis sets is a successful approach, here we take an alternative route which does not require finding new basis sets. Instead, we show in detail how to address the partitioning problem while keeping the use of the original non-orthogonal basis sets on which the DFT code of interest is based on. Partitioning techniques usually rely on Green's functions formalism. We show here how to obtain effective Hamiltonians from reduced Green's functions by proper projection and reversal engineering operations in a general non-orthogonal metric. We calculate the projected or local density of states (LDOS) and projected or local charge in simple examples to illustrate the differences and  the arbitrariness inherent to different ways of performing the \textit{ a priori} same physical partition. The conductance, instead, does not depend on the type of projection, as long as this removes parts of the electrodes which are sufficiently far away from each other. We also discuss, for completeness, the use of mixed basis set where direct and dual subspaces are employed\citep{Thygesen:prb:06}. 

While this work is mainly motivated by problems where standard DFT fails, the partitioning technique presented here can also be used in a variety of other situations such as constrained DFT methodologies\cite{Wu06, Dederichs:PRL:84, PhysRevB.75.115409} or when rigorous population analysis are needed. Some of the mathematical and conceptual issues discussed here have been previously addressed in the literature \citep{Dyall:JPB:84,stechel:JCP:84,Reuter:JCP:13,jacob:prb:10b,
priyadarshy:JCP:96,Ryndyk:arxiv:12, Artacho:PRA:91, Ballentine:86,PhysRevB.83.245124}, but we believe that our approach to the problem may shed a new light on some of the unresolved or controversial points. We begin by revisiting some basic operator operations in a non-orthogonal metric in Sec. \ref{basics}. In Sec. \ref{dual} we discuss the dual basis and its relevance to the inversion operation. Section \ref{basicprojections} presents the essential definitions of projectors along with the different projections that can be carried out and in Sec. \ref{projections}   the concept of non-integer dimension of a subspace is introduced. Section \ref{section_block} revisits the block-orthogonal metrics introduced by Thygesen \citep{Thygesen:prb:06} along with the appropriate  basis change transformations. In Sec. \ref{projected_charge} we show how different projections of the Green's function give rise to different definitions of the local density of states (LDOS) and associated integrated charges, the popular Mulliken charges being one of them. In Sec. \ref{reduced_green} we show how to carry out reversal engineering procedures to obtain effective Hamiltonians from the projected Green's function and, finally, in Sec. \ref{conductance} we show how the conductance can be evaluated starting from any type of projection, being the result independent on the chosen one. A brief set of conclusions is presented at the end.

\section{Operators in an arbitrary metric}
\label{basics}
\noindent We consider a vector space expanded  by a finite and not necessarily orthogonal basis set $\{\ketq i \ket \}$ of dimension $N$. The inner product of its elements $\bra i | j \ket = S_{ij}$ constitutes the overlap matrix, also called the \textit{metric} of such basis set. We will assume, for simplicity, $\{ \ketq i \ket \}$ to be real and   normalized ($\bra i | i \ket = 1  \; \forall \, i$) so that $\bfm{S}$ is real symmetric. For the particular case of an orthogonal basis set the overlap matrix becomes the identity matrix $S_{ij}=\delta_{ij}$. 
The identity operator in terms of this generic basis set is given by the completeness (or closure) relation: 
\begin{equation}
 \bfm{\hat{I}}
 =
 \sum_{ij}{
  \ketq i \ket S^{-1}_{ij} \bra j \braq
 }.
 \label{EQ0001}
 \end{equation}
\noindent The representation of the identity operator  in the basis set defining the metric is  the overlap matrix itself, 
$\bra m \braq  \bfm{\hat{I}} \ketq n \ket = S_{mn} \equiv \left( \bfm{S}\right)_{mn} $. For simplicity, we will write $S^{-1}_{ij}$ instead of the less ambiguous expression  $\left(\bfm{S}^{-1}\right)_{ij}$ for the elements of the inverse matrix. (Unless deemed necessary, this will apply to any matrix inversion from now on). \\
In general, any one-body operator $\bfm{\hat{A}}$ can be written or expressed in terms of the basis set defining the metric:
\bequ 
 \bfm{\hat{A} }  =
 \sum_{ij}{ 
  \ketq  i \ket \tilde{A}_{ij} \bra j \braq
 },
 \label{EQ0002}
\eequ
where 
$\bfm{\tilde{A}}= \bfm{S^{-1}AS^{-1}}$ is 
 the ``nucleus'' or ``nuclear'' matrix of the operator.
One can easily obtain the matrix elements of the operator from the previous expression:
\beque
\begin{split} 
\bra m \braq  \bfm{\hat{A}} \ketq n \ket  = &
 \sum_{ij}{ 
 \bra m |  i \ket \tilde{A}_{ij} \bra j | n \ket
 } = \\
&
 \sum_{ij}{ 
 S_{mi} \tilde{A}_{ij} S_{jn}
 } = A_{mn} 
. 
\end{split}
\eeque
Based on the generalized expression for an operator introduced in Eq. \ref{EQ0002} one can check a basic property for the identity operator:
\beque
\begin{split}
 \bfm{\hat{I}\hat{A}} 
 = &
 \sum_{i j k l}{
  \ketq i \ket S^{-1}_{ij} \bra j | k \ket \tilde{A}_{kl} \bra l \braq
 } = 
  \sum_{i j k l}{
  \ketq i \ket S^{-1}_{ij} S_{jk} \tilde{A}_{kl} \bra l \braq
 } =
\\ &
  \sum_{i k l}{
  \ketq i \ket \delta_{ik} \tilde{A}_{kl} \bra l \braq
 } = 
  \sum_{i l}{
  \ketq i \ket \tilde{A}_{il} \bra l \braq
 } =
 \bfm{\hat{A}}.
 \end{split}
\eeque
More interesting is the following result for the product of two
operators \citep{Dyall:JPB:84}:
\beque
 \bfm{\hat{B}\hat{A}} 
 = 
 \sum_{i j k l}{
  \ketq i \ket \tilde{B}_{ij} \bra j | k \ket \tilde{A}_{kl} \bra l \braq
 }  = 
   \sum_{i j k l}{
  \ketq i \ket \tilde{B}_{ij} S_{jk} \tilde{A}_{kl} \bra l \braq
 },  
\eeque
expression from which the matrix elements of the product can be easily obtained
\bequ
\begin{split}
  \bra m \braq  \bfm{\hat{B}\hat{A}} \ketq n \ket
&=
 \sum_{ijkl}{ 
 \bra m |  i \ket \tilde{B}_{il} \bra l | k \ket \tilde{A}_{kj} \bra j | n \ket
 } = 
 \\
 &
 \sum_{ijkl}{ 
 S_{mi} \tilde{B}_{il} S_{lk} \tilde{A}_{kj} S_{jn} 
 } = 
 \left(  \bfm{BS^{-1}A} \right)_{mn} .
 \end{split}
 \label{EQ0003}
 \eequ
This result can be easily overlooked. In a general metric the product of operators is not simply the product of their matrix  representations. 

\section{The dual basis}
\label{dual}
\noindent Given a general basis set with a metric $\bfm{S}$ (hereon called the \emph{direct} basis set), there
exists a basis set \emph{dual} to this with the property $\bra i | j^{*} \ket = \delta_{ij}$. Only for orthogonal basis sets both direct and dual sets coincide. In general, the nuclear matrix of any operator, $\bfm{\tilde{A}}$, is the dual representation of this operator:
\beque
\bra k^{*} \braq \bfm{\hat{A} } \ketq  l^{*} \ket =
 \bra k^{*} \braq
 \left( \sum_{ij}{ 
  \ketq i \ket \tilde{A}_{ij} \bra j \braq  
 } \right) \ketq l^{*} \ket = 
 \tilde{A}_{kl}.
\eeque
One can equally write or express any operator in terms of the dual basis set:
\beque
 \bfm{\hat{A} }  =
 \sum_{ij}{ 
  \ketq  i^* \ket A_{ij} \bra j^* \braq
 },
 \eeque
where now the nuclear matrix is the representation of the operator in the direct basis itself $\bfm{A}$. This
can be easily proved by direct computation of the matrix elements or by noting that 
\bequ
 \ketq  i \ket = \sum_{j} S_{ij}\ketq j^* \ket, 
  \label{EQ0005}
\eequ
or equivalently  
\bequ
 \ketq  i^* \ket = \sum_{j} S_{ij}^{-1}\ketq j \ket .
  \label{EQ0006}
\eequ
As an example, the identity operator can also be written as
\beque
 \bfm{\hat{I}}
 = 
 \sum_{ij}{
  \ketq i^{*} \ket S_{ij} \bra j^{*} \braq
 }
\eeque
and, for later use, one can also write the identity operator in the following ``orthogonal'' form:
\beque
 \bfm{\hat{I}}
 = 
  \sum_{i}{ 
 \ketq  i^{*} \ket  \bra i \braq
 } =
 \sum_{i}{
  \ketq i \ket \bra i^{*} \braq
 }.
 \eeque
An interesting use of the dual basis  concerns the inversion of an operator. The inversion operation consists of the inversion of the nuclear matrix along with the ``dualization'' of the basis in which this operator is written: 
\bequ
 \bfm{\hat{A}}^{-1} 
 =
 \left[
 \sum_{ij}{ 
 \ketq  i \ket \tilde{A}_{ij} \bra j \braq
 } \right]^{-1} = 
 \sum_{ij}{ 
 \ketq  j^{\ast} \ket\tilde{A}^{-1}_{ji} \bra i^{\ast} \braq
 }. 
\label{EQ0007}
\eequ
The validity of this expression can easily be checked by direct verification of $\bfm{\hat{A}}^{-1} \bfm{\hat{A}}=\bfm{\hat{I}}$. The resulting matrix elements of the inverse operator thus become
\beque
\begin{split}
\bra m \braq  \bfm{\hat{A}}^{-1} \ketq n \ket 
 = &
 \sum_{ij}{ 
 \bra m |  j^{\ast}\ket \tilde{A}^{-1}_{ji} \bra i^{\ast} | n \ket
 } = \\
 &
 \sum_{ij}{ 
 \delta_{mj} \tilde{A}^{-1}_{ji} \delta_{in}
 } = 
  \tilde{A}^{-1}_{mn}. 
 \end{split}
\eeque
Note that the inversion of an operator does not correspond to the inversion of its matrix representation, but to the inversion of its nuclear matrix. 

\section{Projection Operators: Basic considerations}
\label{basicprojections}
\noindent For future use we now establish the basics concerning projection operations. We begin by considering in this section the expression of the operator for the  projection  onto a one-dimensional vector subspace expanded by a single element $\ketq i \ket$ of the direct basis set. Taking into account the definition of the identity operator in Eq. \ref{EQ0001}, it follows
\bequ
  \bfm{\hat{P}}_{i} 
 =
\sum_{j}{
  \ketq i \ket S^{-1}_{ij} \bra j \braq
 }  = \ketq i \ket \bra i^* \braq = \sum_{j}{
  \ketq j^{*} \ket S_{ji} \bra i^{*} \braq
},  \label{EQ0008}
\eequ
where we have used the transformation between direct and dual basis, given in Eqs. \ref{EQ0005} and \ref{EQ0006}, to obtain the three equivalent expressions. With a non-orthogonal basis set one can also write the adjoint of the previous operator:
\bequ
 \bfm{\hat{P}}_{i}^\dagger
 = 
 \sum_{j}{
  \ketq j \ket S^{-1}_{ji} \bra i \braq
 }= \ketq i^* \ket \bra i \braq = \sum_{j}{
  \ketq i^{*} \ket S_{ij} \bra j^{*} \braq }.
 \label{EQ0009}
\eequ
It is easy to show with 
the help of a generic element of the basis set, $\ketq k \ket$, that the projector operator is non-Hermitian by comparing
\beque
 \bfm{\hat{P}}_{i} \ketq k \ket = \sum_{j}{
  \ketq i \ket S^{-1}_{ij} \bra j 
  | k \ket} =
  \delta_{ik} \ketq i \ket 
\eeque
with
\beque
 \bra k \braq \bfm{\hat{P}}_{i} = \sum_{j}{
  \bra k | i \ket S^{-1}_{ij} \bra j \braq
  } =
  \bra i^{*} \braq S_{ik}, 
\eeque
which are obviously not the dual of each other. 
The Hermitian-adjoint projector behaves similarly:
\beque
 \bra k \braq \bfm{\hat{P}}^{\dagger}_{i} = \sum_{j}{
  \bra k | j \ket S^{-1}_{ji} \bra i \braq
  } =
  \bra i \braq \delta_{ik} 
\eeque
and
\beque
 \bfm{\hat{P}}^{\dagger}_{i} \ketq k \ket = \sum_{j}{
  \ketq j \ket S^{-1}_{ji} \bra i 
  | k \ket} =
  S_{ik} \ketq i^{*} \ket 
\eeque
are not dual of each other either. 
This result has important implications when we intend to project a generic operator $\hat{\bfm{A}}$ onto the mono-dimensional  vector subspace expanded by $\ketq i \ket$. One can perform what we call a  Hermitian projection using both projectors  (Eq. \ref{EQ0008} and \ref{EQ0009}):
\beque
\begin{split}
\bfm{\hat{A}}_i\equiv & 
\bfm{\hat{P}}_{i}^{\dagger} \bfm{\hat{A}} \bfm{\hat{P}}_{i} 
= 
\sum_{kl}{
\ketq i^{*} \ket  \bra i | k \ket \tilde{A}_{kl} \bra l | i \ket \bra i^{*} \braq
} 
= \\ 
&
\ketq i^{*} \ket A_{ii} \bra i^{*} \braq. 
\end{split}
\eeque
Representing the resulting operator in the direct basis element one obtains
\beque
\bra i \braq \bfm{\hat{P}}^{\dagger}_{i} \bfm{\hat{A}} \bfm{\hat{P}}_{i}
\ketq i \ket =
\bra i | i^{*} \ket A_{ii} \bra i^{*} | i \ket
 = A_{ii}. 
\eeque
This projection has taken us  onto the corresponding  subspace expanded by the dual basis element
which gives the matrix element $A_{ii}$ as the representation in the direct basis element. 
There is an obvious alternative way of performing the above projection operation:
\beque
\begin{split}
\bfm{\hat{A}}_{i}^* \equiv  & \bfm{\hat{P}}_{i} \bfm{\hat{A}} \bfm{\hat{P}}_{i}^\dagger 
= 
\sum_{jklm}{
\ketq i \ket \bra i^{*} | k \ket \tilde{A}_{kl} \bra l | i^{*} \ket \bra i \braq
}
= \\
&
\sum_{klm}{
\ketq i \ket \delta_{ik} \tilde{A}_{kl} \delta_{li} \bra i \braq
}
= 
\ketq i \ket \tilde{A}_{ii} \bra i \braq. 
\end{split}
\eeque
Representing the resulting operator on the direct basis element one obtains
\beque
\bra i \braq \bfm{\hat{P}}_{i} \bfm{\hat{A}} \bfm{\hat{P}}_{i}^{\dagger}
\ketq i \ket =
\bra i | i \ket \tilde{A}_{ii} \bra i | i \ket
 = \tilde{A}_{ii}. 
\eeque
Notice that the representation of this projection gives the corresponding element of the nuclear matrix and not of the matrix representation of the operator, $A_{ii}$. We call this a \textit{dual} projection as opposed to the previous \textit{direct} projection. \\
For future reference note that one can perform a non-Hermitian projection as
\beque
\begin{split}
&
\bfm{\hat{P}}_{i} \bfm{\hat{A}} \bfm{\hat{P}}_{i} 
= 
\sum_{kl}{
\ketq i \ket \bra i^{*} | k \ket \tilde{A}_{kl} \bra l | i \ket \bra i^{*} \braq
}
= \\
&
\sum_{klm}{
\ketq i \ket \delta_{ik} \tilde{A}_{kl} S_{li} S^{-1}_{im} \bra m \braq
}
= 
\sum_{lm}{
\ketq i \ket \tilde{A}_{il} S_{li} S^{-1}_{im} \bra m \braq
},
\end{split}
\eeque
from which, when represented on the mono-dimensional direct subspace, one obtains
\bequ
\bra i \braq \bfm{\hat{P}}_{i} \bfm{\hat{A}} \bfm{\hat{P}}_{i} \ketq i \ket = 
\sum_{lm}{
\bra i | i \ket \tilde{A}_{il}  S_{li} S^{-1}_{im} \bra m | i \ket
}
= 
\left(\bfm{\tilde{A}S}\right)_i.
\label{PP}
\eequ
Note also that there is an alternative way of performing a non-Hermitian projection
\beque
\begin{split}
&\bfm{\hat{P}}^{\dagger}_{i} \bfm{\hat{A}} \bfm{\hat{P}}^{\dagger}_{i} 
= 
\sum_{kl}{
\ketq i^{*} \ket \bra i | k \ket \tilde{A}_{kl} \bra l | i^{*} \ket \bra i \braq
}
= \\
&
\sum_{klm}{
\ketq m \ket S^{-1}_{mi} S_{ik}  \tilde{A}_{kl}  \delta_{li} \bra i \braq
}
\end{split}
\eeque
with an associated representation given by
\bequ
 \left(\bfm{S\tilde{A}}\right)_i.
 \label{P*P*}
\eequ
Finally, and also for future reference, a straightforward application of the inversion operation (Eq. \ref{EQ0007}) onto the projector operator gives
\beque
\begin{split}
\left[\bfm{\hat{P}}^{\dagger}_{i} \right]^{-1}  = 
\left[\sum_{j}{
  \ketq j \ket S^{-1}_{ji} \bra i \braq
 } \right]^{-1} = 
\sum_{j}{
  \ketq i^{*} \ket S_{ij} \bra j^{*} \braq
 }= 
\bfm{\hat{P}}^{\dagger}_{i}.
\end{split}
\eeque
Since $\left[\bfm{\hat{P}}^\dagger_{i} \right]^{-1}\bfm{\hat{P}}^\dagger_{i} = \bfm{\hat{P}}^\dagger_{i}\bfm{\hat{P}}^\dagger_{i} \neq \bfm{\hat{I}}$, the inversion of a projector operator is not a true inverse, but a generalized inverse which obeys the Penrose condition \citep{Penrose:55}:
\beque
 \bfm{\hat{P}}^{\dagger}_{i}\left[\bfm{\hat{P}}^{\dagger}_{i} \right]^{-1}\bfm{\hat{P}}^{\dagger}_{i} =\bfm{\hat{P}}^{\dagger}_{i}.
\eeque

\section{Projection onto a subspace: Integer and non-integer dimension.}
\label{projections}
\begin{figure}[t]
    		    \includegraphics*[width=0.30\textwidth]{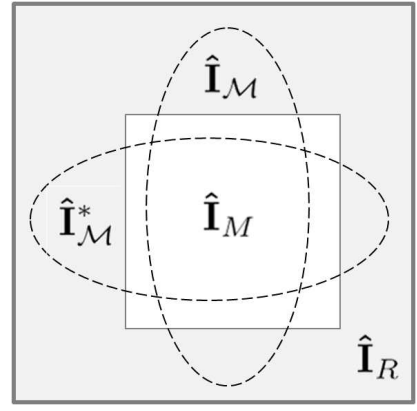}   
  \caption{
    Schematic representation of the two types of projections explained in the text and the resulting subspaces. Areas within the dashed lines represent subspaces with non-integer dimension, i.e., not orthogonal to the remaining area or subspace. The areas delimited by the solid line represent subspaces with integer dimensions, i.e., orthogonal to each other.
   \label{FIG01}}
\end{figure}

\noindent The projection onto a subspace of dimension bigger than one is now carried out by the generalized projectors
\beque
\begin{split}
 \bfm{\hat{P}}_{\rm M} &
 =
  \sum_{m , i}{
   \ketq m \ket S^{-1}_{m i} \bra i \braq
 } =  \sum_{m}{
   \ketq m \ket  \bra m^* \braq
 } , \\
 \bfm{\hat{P}^ \dagger}_{\rm M} &
 =
  \sum_{m, i}{
   \ketq i \ket S^{-1}_{i m} \bra m \braq
 } = \sum_{m}{
   \ketq m^* \ket \bra m \braq
 } , 
 \end{split}
\eeque
where $m$ runs now over a selected subset M  consisting of $N_{\rm M} (< N)$ elements of the direct basis set. As can be seen in the right-hand expressions, this restricted summation also implies to run over the corresponding subset in the dual space. The letters $i$ and $j$ will always denote the elements of the full direct and dual spaces from now on.  The remaining direct basis elements  constitute the subset R such that 
\beque
 \bfm{\hat{I}}=
 \bfm{\hat{P}}_{\rm M} + \bfm{\hat{P}}_{\rm R} =
 \bfm{\hat{P}}^{\dagger}_{\rm M} +\bfm{\hat{P}}^{\dagger}_{\rm R}. 
\eeque
We now propose to write our full vector space as
\beque
\bfm{\hat{I}}= (\bfm{\hat{P}}_{\rm M} + \bfm{\hat{P}}_{\rm R}) \bfm{\hat{I}} (\bfm{\hat{P}}_{\rm M} + \bfm{\hat{P}}_{\rm R})  
\\
\bfm{\hat{I}}= (\bfm{\hat{P}}^{\dagger}_{\rm M} + \bfm{\hat{P}}^{\dagger}_{\rm R}) \bfm{\hat{I}} (\bfm{\hat{P}}^{\dagger}_{\rm M} + \bfm{\hat{P}}^{\dagger}_{\rm R}) 
\\
\bfm{\hat{I}}= (\bfm{\hat{P}}_{\rm M} + \bfm{\hat{P}}_{\rm R}) \bfm{\hat{I}} (\bfm{\hat{P}}^{\dagger}_{\rm M} + \bfm{\hat{P}}^{\dagger}_{\rm R})
\\
\bfm{\hat{I}}= (\bfm{\hat{P}}^{\dagger}_{\rm M} + \bfm{\hat{P}}^{\dagger}_{\rm R}) \bfm{\hat{I}} (\bfm{\hat{P}}_{\rm M} + \bfm{\hat{P}}_{\rm R})
\eeque
The four equivalent expressions give rise to four different ways
of partitioning the full vector space:
\begin{eqnarray}
\bfm{\hat{I}}&=& 
\bfm{\hat{P}}_{\rm M} \bfm{\hat{I}}\bfm{\hat{P}}_{\rm M} 
 +\bfm{\hat{P}}_{\rm R}\bfm{\hat{I}}\bfm{\hat{P}}_{\rm R}
 \label{proper1} 
\\
\bfm{\hat{I}}&=& 
\bfm{\hat{P}}^{\dagger}_{\rm M} \bfm{\hat{I}}\bfm{\hat{P}}^{\dagger}_{\rm M} 
 +\bfm{\hat{P}}^{\dagger}_{\rm R}\bfm{\hat{I}}\bfm{\hat{P}}^{\dagger}_{\rm R} 
\label{proper2}
\\
\bfm{\hat{I}}&=& 
\bfm{\hat{P}}_{\rm M} \bfm{\hat{I}}\bfm{\hat{P}}^{\dagger}_{\rm M} + \bfm{\hat{P}}_{\rm M}\bfm{\hat{I}}\bfm{\hat{P}}^{\dagger}_{\rm R}+\bfm{\hat{P}}_{\rm R}
\bfm{\hat{I}}\bfm{\hat{P}}^{\dagger}_{\rm M} +\bfm{\hat{P}}_{\rm R}\bfm{\hat{I}}\bfm{\hat{P}}^{\dagger}_{\rm R} 
\label{proper3}
\\
\bfm{\hat{I}}&=& 
\bfm{\hat{P}}^{\dagger}_{\rm M} \bfm{\hat{I}}\bfm{\hat{P}}_{\rm M} + \bfm{\hat{P}}^{\dagger}_{\rm M}\bfm{\hat{I}}\bfm{\hat{P}}_{\rm R}+\bfm{\hat{P}}^{\dagger}_{\rm R}
\bfm{\hat{I}}\bfm{\hat{P}}_{\rm M} +\bfm{\hat{P}}^{\dagger}_{\rm R}\bfm{\hat{I}}\bfm{\hat{P}}_{\rm R} 
\label{proper4}
\end{eqnarray}
By leaving $\bfm{\hat{I}}$ in between the projectors the meaning of each term as partial projections of the original full vector space becomes clear. Equations \ref{proper1} and \ref{proper2} contain what we have called non-Hermitian projections in previous section. There the full vector space has been partitioned into two complementary orthogonal subspaces (thereby the two cross projections are missing in the expressions). Equations \ref{proper3} and \ref{proper4} contain Hermitian projections. These, as we explain below, do not generate proper orthogonal vector spaces since the cross projections do not vanish.
A graphical illustration of the different projection operations which will be useful in the ensuing discussion is shown in Fig. \ref{FIG01}. 

We will first consider the result of performing the direct Hermitian projection:
\beque
\begin{split} 
\bfm{\hat{I}}_{\mathcal{M}} = \bfm{\hat{P}}^{\dagger}_{\rm M} \bfm{\hat{I}} \bfm{\hat{P}}_{\rm M} = &
\sum_{m, n} {\ketq m^{*} \ket \bra m \braq \bfm{\hat{I}} \ketq n \ket \bra n^{*} \braq }= 
\\
&
\sum_{m,n}{ 
\ketq m^{*} \ket S_{mn} \bra n^{*} \braq}. 
\end{split}
\eeque
The representation of this projection in the direct basis set is given by 
\beque
\bra m \braq  \bfm{\hat{I}}_{\mathcal{M}} \ketq n \ket =
S_{mn}\equiv \bfm{S}_{\rm M}.
\eeque
While this result was somewhat expected, it hides a pitfall  since $\bfm{\hat{I}}_{\mathcal{M}}$ is not idempotent (this is obvious by noticing that $\bfm{\hat{I}}_{\mathcal{M}} \bfm{\hat{I}}_{\mathcal{M}} \ne \bfm{\hat{I}}_{\mathcal{M}}$). Therefore the partition it represents is not a proper vector subspace. This prompt us to introduce a quantity defined by the integration of the diagonal elements of a given projection represented in real space.  For the full vector space we started with this one writes
\beque
D = \int{\bra \vec{r} \braq \bfm{\hat{I}} \ketq \vec{r} \ket d\vec{r}}  .
\eeque
Explicitly evaluated  one obtains
\beque
\begin{split}
D =&
 \int{
\sum_{i,j}{
\bra \vec{r} | i^{*} \ket S_{ij} \bra j^{*} | \vec{r} \ket }d\vec{r}} = 
\sum_{i,j}{
S_{ij} \int{ \phi^{*}_{i} \phi^{*\dagger}_{j}d\vec{r}}} = 
\\
&
\sum_{i,j}{
S_{ij} S^{-1}_{ji}} = Tr \left[\bfm{S}\bfm{S^{-1}}\right] = Tr \left[\bfm{I}\right] = N,
\end{split}
\eeque
which, obviously, is the number of elements of the full starting space, $N$. On the other hand, when one evaluates such quantity for $\bfm{\hat{I}}_{\mathcal{M}}$ one obtains  
\beque
D_{\mathcal{M}} = \int{\bra \vec{r} \braq\bfm{\hat{I}}_{\mathcal{M}}  \ketq \vec{r} \ket d\vec{r}}  = Tr \left[\bfm{S}_{\rm M}\bfm{S}^{-1}_{\rm M} \right],
\eeque
which is clearly $\ne N_{\rm M}$ and may be a non-integer value. This ``subspace'' with non-integer dimension is represented in Fig. \ref{FIG01} by the oval area within the dashed lines. Another manifestation of this non-integer dimension is given by the representation of $\bfm{\hat{I}}_{\mathcal{M}}$ in the dual subspace
\beque
\bra m^{*}\braq \bfm{\hat{I}}_{\mathcal{M}}\ketq n^{*} \ket = 
\bfm{S}^{-1}_{\rm M}\bfm{S}_{\rm M}\bfm{S}^{-1}_{\rm M}.
\eeque
which is not equal to $\bfm{S}^{-1}_{\rm M}$, as naively expected. 

The generalization of the direct Hermitian projection to any operator $\bfm{\hat{A}}$ is given by
\bequ
\bfm{\hat{A}}_{\mathcal{M}}=\bfm{\hat{P}}^{\dagger}_{\rm M}\bfm{\hat{A}}\bfm{\hat{P}}_{\rm M} = \sum_{m,n} 
\ketq m^* \ket A_{mn} \bra n^* \braq,
\label{P*P}
\eequ
whose representation in the direct basis subset is 
\beque
\bra m \braq \bfm{\hat{A}}_{\mathcal{M}}
\ketq n \ket = 
 \bfm{A}_{\rm M}
\eeque
and in the dual basis subset is
\beque
\bra m^{*}\braq \bfm{\hat{A}}_{\mathcal{M}} \ketq n^* \ket  = 
 \bfm{S}^{-1}_{\rm M}\bfm{A}_{\rm M}\bfm{S}^{-1}_{\rm M}. 
\eeque
Likewise, the dual Hermitian projection, already introduced in the one-dimensional case, defines a new ``subspace''
 \beque
 \begin{split}
\bfm{\hat{I}}^{*}_{\mathcal{M}}= \bfm{\hat{P}}_{\rm M} \bfm{\hat{I}} \bfm{\hat{P}}^{\dagger}_{\rm M} = & \sum_{m,n} {\ketq m \ket \bra m^{*} \braq \bfm{\hat{I}} \ketq n^{*} \ket \bra n \braq } =  \\
&
\sum_{m,n}{ 
\ketq m \ket S^{-1}_{mn} \bra n \braq}
\end{split}
\label{Id_M}
\eeque  
with a non-integer dimension given by
\beque
D^*_{\mathcal{M}} = \int{\bra \vec{r} \braq \bfm{\hat{I}}^{*}_{\mathcal{M}} \ketq \vec{r} \ket d\vec{r}}  = Tr \left[\bfm{S}^{-1}_{\rm M}\bfm{S}_{\rm M} \right]. 
\eeque
which coincides with $D_{\mathcal{M}}$.
Figure \ref{FIG01} depicts this subspace as a different oval area within dashed lines. Its
representation in the dual subspace is the ``expected'' one
\beque
\bra m^{*} \braq \bfm{\hat{I}}^{*}_{\mathcal{M}}
\ketq n^{*} \ket = 
\bfm{S}^{-1}_{\rm M},
\eeque 
but its representation in the direct subspace is not
\bequ
\bra m \braq \bfm{\hat{I}}^{*}_{\mathcal{M}}
\ketq n \ket = 
 \bfm{S}_{\rm M}\bfm{S}^{-1}_{\rm M}\bfm{S}_{\rm M}.
\label{unorthodox}
\eequ

The generalized dual Hermitian projection to any operator $\bfm{\hat{A}}$ is thus given by
\bequ
\bfm{\hat{A}}^{*}_{\mathcal{M}}=\bfm{\hat{P}}_{\rm M} \bfm{\hat{A}} \bfm{\hat{P}}_{\rm M}^\dagger = 
\sum_{m,n} {
\ketq m \ket \tilde{A}_{mn} \bra n \braq},
\label{dualprojection}
\eequ
whose representation in the direct basis subset is 
\beque
\bra m \braq \bfm{\hat{A}}^{*}_{\mathcal{M}}
\ketq n \ket = 
\bfm{S}_{\rm M}\bfm{\tilde{A}}_{\rm M}\bfm{S}_{\rm M}.
\eeque
Notice that this matrix multiplication does not only involve matrices of reduced dimension since the evaluation of $ \bfm{\tilde{A}}$ requires matrix representations  in the initial full vector space. 

Lastly and for completeness we note that we can also perform non-Hermitian projections. These projections also generate new subspaces, which in this case have integer dimensions:
\beque
\begin{split}
D_{\rm M} = &
\int{\bra \vec{r} \braq \bfm{\hat{P}}^{\dagger}_{\rm M} \bfm{\hat{I}} \bfm{\hat{P}}^{\dagger}_{\rm M} \ketq \vec{r} \ket d\vec{r}} = 
\int{\bra \vec{r} \braq \bfm{\hat{P}}_{\rm M} \bfm{\hat{I}} \bfm{\hat{P}}_{\rm M} \ketq \vec{r} \ket d\vec{r}}  = \\
& = \int{\bra \vec{r} \braq  \bfm{\hat{I}}_{\rm M} \ketq \vec{r} \ket d\vec{r}}  = Tr \left[\bfm{I}_{\rm M} \right] = N_{\rm M}.
\end{split}
\eeque
The subspace represented by $\bfm{\hat{I}}_{\rm M}$ is now orthogonal to
the rest of the initial vector space. This is depicted by the square area within solid lines in Fig. \ref{FIG01}.
Operators projected in this manner can also be represented on the direct and the dual subspaces and we will make use of these representations in the following sections. It is, however, worth noting here that in order to recover the equivalent of the one-dimensional representations in Eqs. \ref{PP} and \ref{P*P*}
one must represent at the same time on both  direct and  dual subsets which are orthogonal to each other: 
\beque
\bra m^{*} \braq \bfm{\hat{P}}_{\rm M}\bfm{\hat{A}} \bfm{\hat{P}}_{\rm M} \ketq n \ket =
 \left(\bfm{\tilde{A}S}\right)_{\rm M} 
\eeque
and similarly: 
\beque
\bra m \braq \bfm{\hat{P}}^{\dagger}_{\rm M}\bfm{\hat{A}} \bfm{\hat{P}}^{\dagger}_{\rm M} \ketq n^{*} \ket =
 \left(\bfm{S\tilde{A}}\right)_{\rm M}.
\eeque 

\section{Block orthogonal metrics.}
\label{section_block}
\noindent We revisit now, from a projection perspective, block-orthogonal metrics which were already introduced in Ref. \onlinecite{Thygesen:prb:06}. The projectors introduced in previous section can be used to build a new basis set:
\beque
\begin{split}
&\left\{ \ketq i_{\Delta} \ket \right\} = \bfm{\hat{P}}_{\rm M}\{ \ketq i \ket \} + \bfm{\hat{P}}^{\dagger}_{\rm R} \{ \ketq i^{*} \ket \} = \\
&
\{\sum_{m,j,i}{ \ketq m \ket S^{-1}_{mj}\bra j | i \ket }\} + \{ \sum_{r,j,i}{ \ketq r^{*} \ket S_{rj}\bra j^{*} | i^{*} \ket \} }= \\
&
\{\ketq m \ket \} + \{ \ketq r^{*} \ket \} .
\end{split}
\eeque
\beque
\begin{split}
&\left\{ \ketq i_{\nabla} \ket \right\} = \bfm{\hat{P}}^{\dagger}_{\rm M}\{ \ketq i^{*} \ket \} + \bfm{\hat{P}}_{\rm R} \{ \ketq i \ket \} = \\
&
\{\sum_{m,j,i}{ \ketq m^{*} \ket S_{mj}\bra j^{*} | i^{*} \ket }\} + \{ \sum_{r,j,i}{ \ketq r \ket S^{-1}_{rj}\bra j | i \ket \} }= \\
&
 \{ \ketq m^{*} \ket \} + \{\ketq r \ket \}.
\end{split}
\eeque
Here the labels $m $ and $r$ run only over the subspace selected by the projector operators. 
The representation of the identity operator in the basis set $\left\{ \ketq i_{\nabla} \ket \right\}$ is given by
\beque
 \bra m \braq \bfm{\hat{I}} \ketq n \ket = 
\sum_{i,j}\bra m | i\ket S^{-1}_{ij}\bra j | n \ket = \bfm{S}_{\rm M}
\eeque
\bequ 
 \bra r^* \braq \bfm{\hat{I}} \ketq s^* \ket = 
\sum_{i,j}\bra r^* | i\ket S^{-1}_{ij}\bra j | s^* \ket = \bfm{S}_{\rm R}^{-1}
\nonumber
\eequ
\bequ 
 \bra m \braq \bfm{\hat{I}} \ketq r^* \ket = 
\sum_{i,j}\bra m | i\ket S^{-1}_{ij}\bra j | r^* \ket  = 0
\nonumber
\eequ
\bequ 
 \bra r^* \braq \bfm{\hat{I}} \ketq m \ket = 
\sum_{i,j}\bra r^* | i\ket S^{-1}_{ij}\bra j | m \ket  = 0
\nonumber
\eequ
and similarly for the basis set $\left\{ \ketq i_{\nabla} \ket \right\} $. Here the labels $m,n$ and $r, s$ run only over the subspaces M and R$^*$, respectively. In matrix form the block-orthogonality becomes clear: 
\bequ
\bfm{S}_{\Delta} = \left(\begin{array}{cc}
      \bfm{S}_{\rm M} & 0  \\
      0 & \bfm{S}^{-1}_{\rm R} 
   \end{array}\right)
,\hspace{0.5ex} 
\bfm{S}_{\nabla} = \left(\begin{array}{cc}
    \bfm{S}^{-1}_{\rm M} & 0  \\
    0 & \bfm{S}_{\rm R} 
   \end{array}\right).
   \label{block}
\eequ
The full identity operator can now be re-expressed with the help of these new basis sets as
\beque
\begin{split}
 \bfm{\hat{I}} & =   
\sum_{i,j}{\ketq i_{\Delta}\ket \left(\bfm{\tilde{S}}_{\Delta}\right)_{ij}\bra i_{\Delta}\braq} = \\
&
\sum_{m,n}{
\ketq m \ket \left(\bfm{\tilde{S}}_{\Delta}\right)_{mn} \bra n \braq}
+ \sum_{m,r}{
\ketq m \ket \left(\bfm{\tilde{S}}_{\Delta}\right)_{mr} \bra r^{*} \braq} + \\
&
\sum_{r,m}{
\ketq r^{*} \ket \left(\bfm{\tilde{S}}_{\Delta}\right)_{rm} \bra m \braq}
+ \sum_{r,s}{
 \ketq r^{*} \ket  \left(\bfm{\tilde{S}}_{\Delta}\right)_{rs}  \bra s^{*} \braq}
\end{split}
\eeque
or
\beque
\begin{split}
 \bfm{\hat{I}} & =   
\sum_{i,j}{\ketq i_{\nabla}\ket \left(\bfm{\tilde{S}}_{\nabla}\right)_{ij}\bra i_{\nabla}\braq} = \\
&
\sum_{m,n}{
 \ketq m^{*} \ket  \left(\bfm{\tilde{S}}_{\nabla}\right)_{mn}  \bra n^{*} \braq}
+ \sum_{m,r}{
\ketq m^{*} \ket  \left(\bfm{\tilde{S}}_{\nabla}\right)_{mr} \bra r \braq} \\
&
+ \sum_{r,m}{
\ketq r \ket \left(\bfm{\tilde{S}}_{\nabla}\right)_{rm}\bra m^{*} \braq}
+ \sum_{r,t}{
\ketq r \ket \left(\bfm{\tilde{S}}_{\nabla}\right)_{rs} \bra s \braq}.
\end{split}
\eeque
The $\bfm{\tilde{S}}_{\Delta}$ and $\bfm{\tilde{S}}_{\nabla}$ matrices are the dual form of the block-orthogonal metrics in Eq. \ref{block} which are explicitly written as
\beque
\bfm{\tilde{S}}_{\Delta} = \left(\begin{array}{cc}
      [\bfm{S}_{\rm M}]^{-1} & 0  \\
      0 & \bfm{S}^{*}_{\rm R} 
   \end{array}\right)
,\hspace{0.5ex} 
\bfm{\tilde{S}}_{\nabla} = \left(\begin{array}{cc}
    \bfm{S}^{*}_{\rm M} & 0  \\
    0 & [\bfm{S}_{\rm R} ]^{-1}
   \end{array}\right).
\eeque
Notice that $\bfm{S}^{*}_{\rm M(R)}$ stands for $[\bfm{S}^{-1}_{\rm M(R)}]^{-1} $. The dual basis in these new metrics has no direct correspondence with the direct and dual original basis sets. 
Using, e.g., the metric $\Delta$, we can now define new projector operators as
\beque
\begin{split}
 \bfm{\hat{P}}_{\rm M}^{\Delta} &
 = \bfm{\hat{P}}_{\rm M}^{\Delta\dagger} =
 \sum_{m,n}{
\ketq m \ket [\bfm{S}_{\rm M}]^{-1} \bra n \braq}  , \\
 \bfm{\hat{P}}^{\Delta}_{\rm R} &
 = \bfm{\hat{P}}_{\rm R}^{\Delta\dagger}=
 \sum_{r,s}{
\ketq r^{*} \ket \bfm{S}^{*}_{\rm R} \bra s^{*} \braq} 
\end{split}
\eeque
which add up to the identity operator
\beque
\bfm{\hat{I}}= 
\bfm{\hat{{P}}}^\Delta_{\rm M} \bfm{\hat{I}}\bfm{\hat{\bfm{P}}}^\Delta_{\rm M} + 
\bfm{\hat{{P}}}^\Delta_{\rm R}\bfm{\hat{I}}
\bfm{\hat{{P}}}^\Delta_{\rm R}. 
\eeque
(In a similar manner one can define projectors associated with the other metric $\nabla$.) The dimension of both partitions is now integer. For instance:
\beque
\begin{split}
D = &
\int{\bra \vec{r} \braq \bfm{\hat{P}}^\Delta_{\rm M} \bfm{\hat{I}} \bfm{\hat{P}}^\Delta_{\rm M} \ketq \vec{r} \ket d\vec{r}} = 
\\
&
\int{
\sum_{m,n}{\bra \vec{r} | m \ket   [\bfm{S}_{\rm M}]^{-1}  \bra n | \vec{r} \ket} d\vec{r}}  = Tr \left[\bfm{I}_{\rm M} \right] = N_{\rm M}.
\end{split}
\eeque
The generalized representation of any one-body operator in terms of these basis sets is given by 
\beque
\bfm{A}_{\Delta} =
\left(\begin{array}{cc}
  \bfm{A}_{\rm M} & (\bfm{\tilde{A}S})_{\rm MR}  \\
  (\bfm{S\tilde{A}})_{\rm RM} & \bfm{\tilde{A}}_{\rm R} 
        \end{array}\right)
\eeque
which is not block-diagonal any more.
The same obviously holds for the other metric $\nabla$:
\beque
				\bfm{A}_{\nabla} =
\left(\begin{array}{cc}
  \bfm{\tilde{A}}_{\rm M} & (\bfm{S\tilde{A}})_{\rm MR}  \\
  (\bfm{\tilde{A}S})_{\rm RM} & \bfm{A}_{\rm R} 
        \end{array}\right).
\label{EQ0027}
\eeque
It is useful to propose matrix transformations connecting all the metrics we have presented so far: 
\beque
\bfm{\Pi}^{\dagger}_{\Delta}\bfm{A}\bfm{\Pi}_{\Delta}=\bfm{A}_{\Delta},
\eeque
and 
\beque
\bfm{\Pi}^{\dagger}_{\nabla}\bfm{\tilde{A}}\bfm{\Pi}_{\nabla}=\bfm{A}_{\nabla}.
\eeque
The matrices $\bfm{\Pi}_{\nabla}$ and $\bfm{\Pi}_{\nabla}$ are defined by
\beque
\bfm{\Pi}_{\Delta} =
\left(\begin{array}{cc}
    \bfm{I} & \bfm{S}^{-1}_{\rm MR}  \\
    0 & \bfm{S}^{-1}_{\rm R} 
       \end{array}\right)
,\hspace{0.5ex}
\bfm{\Pi}_{\nabla} =
\left(\begin{array}{cc}
\bfm{I} & \bfm{S}_{\rm MR}  \\
0 & \bfm{S}_{\rm R} 
        \end{array}\right).
\eeque
The $\bfm{\Pi}$ matrices are overlap matrices between this new basis and the direct basis: $\bra i | i_{\Delta} \ket = (\bfm{\Pi}_{\nabla})_{ii}$ and the dual one $\bra i^{*} | i_{\Delta} \ket = (\bfm{\Pi}_{\Delta})_{ii}$.  We can now write the relationships
\beque
\ketq i \ket = \sum_{j} { \left(\Pi^{-1}_{\nabla}\right)_{ij} \ketq j_{\nabla} \ket }
\hspace{2.0ex}
\ketq i \ket = \sum_{j} {\left( \Pi^{-1}_{\Delta}\right)_{ij} \ketq j_{\Delta} \ket }
\eeque

\section{The projected charge}
\label{projected_charge}

A natural starting point for the evaluation of many physical quantities in infinite systems which are characterized by a one-body Hamiltonian $\bfm{\hat{H}}$ is the Green's function operator: 
\beque  
 \left[
  \left(
 \omega - \mu \pm i\delta
 \right)
 \bfm{\hat{I}} - \bfm{\hat{H}} 
 \right]
 \bfm{\hat{G}}^{(\pm)}(\omega) 
 =
 \bfm{\hat{I}}, 
\eeque
where $\mu$ and $\omega$ 
are the chemical potential and energy, respectively, and +(-) denotes advanced(retarded) Green's function.
For all practical purposes the Hamiltonian must be restricted to a finite region which is connected to the rest of the world through an energy dependent self-energy operator $\bfm{\hat{\Sigma}}(\omega)$:
\bequ  
 \left[
  \left(
 \omega - \mu
 \right)
 \bfm{\hat{I}} - \bfm{\hat{H}} 
  -\bfm{\hat{\Sigma}}(\omega)
 \right]
 \bfm{\hat{G}}(\omega) 
 =
 \bfm{\hat{I}}. 
\label{EQ0029}
\eequ
At this point the Hamiltonian is supposed to represent a large enough system and contain all possible physically relevant information. The particular form of the self-energy operator $\bfm{\hat{\Sigma}}$ (the retarded one from now on) is assumed not to be relevant in what follows.  Making use of Eq. \ref{EQ0003}, one can easily show that the representation of Eq. \ref{EQ0029} in matrix form is
\beque
 \bfm{G}(\omega) =
 \bfm{S}
  \left[
   \left(
 \omega - \mu
 \right)
 \bfm{S} - \bfm{H} - \bfm{\Sigma}(\omega)
 \right]^{-1}
 \bfm{S}.
\eeque
For convenience this is usually written as
\bequ
 \bfm{\tilde{G}}(\omega)
 =
   \left[
 \left(
 \omega - \mu
 \right) \bfm{S}
 - \bfm{H}
 - \bfm{\Sigma}(\omega)
 \right]^{-1},
 \label{gtilde}
\eequ
which, according to Eq. \ref{EQ0007}, can be directly obtained from the matrix representation in the direct basis of the operator equation
\bequ
\left[\bfm{\hat{G}}(\omega)\right]^{-1} = 
    \left(
 \omega - \mu
 \right)
 \bfm{\hat{I}} - \bfm{\hat{H}} 
  - \bfm{\hat{\Sigma}(\omega)}. 
  \label{EQ0032}
\eequ

The partial charge or partial electronic density associated to a part of the space or to a subset of atoms or orbitals\citep{Clark03,Becke88,Fonseca04} (what we have called subspace M in previous sections) can be obtained from integration of the LDOS:  
\bequ
\rho_{\mathcal{M}} = \int^{0}_{-\infty} {D_{\mathcal{M}}(\omega)
d\omega},
\label{charge}
\eequ
which, in turn, is computed through a projection of the Green's function operator. We begin by considering the direct Hermitian projection, as explained in Sec. \ref{projections}. For simplicity's sake, we assume spin degeneracy from now on: 
\beque
D_{\mathcal{M}}(\omega) = -\frac{2}{\pi} Im \left[
\int {
\bra \vec{r} | \bfm{\hat{P}}^{\dagger}_{\rm M} \bfm{\hat{G}(\omega)} \bfm{\hat{P}}_{\rm M} | \vec{r} \ket d\vec{r} }
\right]. 
\eeque
We now carry out the volume integral:
\beque
\begin{split} 
  &\int { 
\bra \vec{r} | \bfm{\hat{P}^{\dagger}}_{\rm M} \bfm{\hat{G}} \bfm{\hat{P}}_{\rm M} | \vec{r} \ket 
d\vec{r}}= \\
 & =\int {
  \sum_{m,n} {
 \bra \vec{r} | m^{*} \ket 
 \bra m | \bfm{\hat{G}} | n \ket \bra n^{*} |\vec{r} \ket } 
 d\vec{r}} = \\
 &
=\sum_{m,n} {
 \bra m | \bfm{\hat{G}} | n \ket \int {\phi^{*}_{m}(\vec{r})\phi^{\dagger *}_{n}(\vec{r})d\vec{r}}} =
  \sum_{m,n} {G_{mn} 
  S_{nm}^{-1} }.
 \end{split}
\eeque
The LDOS associated with this projection is thus given by
\bequ
D_{\mathcal{M}}(\omega) = -\frac{2}{\pi} Im \left\{
  Tr \left[\bfm{G}_{\rm M}(\omega)
  \bfm{S}_{\rm M}^{-1}  \right] \right\}.
	\label{D1}
\eequ
Notice that when one integrates Eq. \ref{charge} up to $\infty$ one gets what we have  previously defined as dimension of the projected subspace $D_\mathcal{M}$. (We will use the same letter for the $\omega$-dependent quantity and the integrated one from now on). 

In the case of the dual Hermitian projection the partial electron density is given by
\beque
\begin{split}
& \rho^{*}_{\mathcal{M}} 
  = -\frac{2}{\pi}\int^{0}_{-\infty} { Im \left[
 \int { 
\bra \vec{r} | \bfm{\hat{P}}_{\rm M} \bfm{\hat{G}(\omega)} \bfm{\hat{P}}^{\dagger}_{\rm M} | \vec{r} \ket 
d\vec{r}} \right]
d\omega } = \\
&
=-\frac{2}{\pi}\int^{0}_{-\infty} { Im \left\{
   Tr \left[
   \bfm{\tilde{G}}_{\rm M}(\omega) \bfm{S}_{\rm M}  \right] \right\} 
   d\omega }.
 \end{split}
\eeque
We therefore may define
\bequ
D^{*}_{\mathcal{M}}(\omega) = -\frac{2}{\pi}  Im \left\{
   Tr \left[
   \bfm{\tilde{G}}_{\rm M}(\omega) \bfm{S}_{\rm M}  \right] \right\}.
\eequ
The non-Hermitian projections give the well-known result known as Mulliken population analysis\cite{mulliken}:
\beque
\begin{split}
& \rho_{\rm M} =  \int^{0}_{-\infty} {D_{\rm M}(\omega) d\omega } = \\
 & = -\frac{2}{\pi}\int^{0}_{-\infty} { Im \left[
 \int { 
\bra \vec{r} | \bfm{\hat{P}}_{\rm M} \bfm{\hat{G}(\omega)} \bfm{\hat{P}}_{\rm M} | \vec{r} \ket 
d\vec{r}} \right]
d\omega } = \\
& = -\frac{2}{\pi}\int^{0}_{-\infty} { Im \left[
 \int { 
\bra \vec{r} | \bfm{\hat{P}}^{\dagger}_{\rm M} \bfm{\hat{G}(\omega)} \bfm{\hat{P}}^{\dagger}_{\rm M} | \vec{r} \ket 
d\vec{r}} \right]
d\omega } = \\
&
=-\frac{2}{\pi}\int^{0}_{-\infty} { Im \left\{
   Tr \left[
   \bfm{\tilde{G}}(\omega) \bfm{S}  \right]_{\rm M} \right\} 
   d\omega }.
\end{split}
\eeque
Finally, one can also evaluate the charge using the block-orthogonal metric $\Delta$ discussed in Sec. \ref{section_block}:
\beque
\rho^{\Delta}_{\rm M} = \int^{0}_{-\infty} { D^{\Delta}_{\rm M}(\omega)
d\omega}.
\eeque
where the LDOS is given by 
\beque
\begin{split}
D_{\rm M}^{\Delta}(\omega) = 
-\frac{2}{\pi} Im \left[
\int {
\bra \vec{r} | \bfm{\hat{P}}^\Delta_{\rm M} \bfm{\hat{G}(\omega)} \bfm{\hat{P}}^\Delta_{\rm M} | \vec{r} \ket d\vec{r} }
\right] .
\end{split}
\eeque
Performing the projection and volume integration one obtains
\beque
\begin{split} 
 & \int { 
\bra \vec{r} | \bfm{\hat{P}}^\Delta_{\rm M} \bfm{\hat{G}} \bfm{\hat{P}}^\Delta_{\rm M} | \vec{r} \ket 
d\vec{r}}= \\
&
  \int {
  \sum_{k,l} {
 \bra \vec{r} | k^{*} \ket 
 \bra k | \bfm{\hat{G}} | l \ket  \bra l^{*} |\vec{r} \ket } 
 d\vec{r}} = \\
 &
\sum_{k,l} {
 \bra k | \bfm{\hat{G}} | l \ket \int {\phi^{*}_{k}(\vec{r})\phi^{\dagger *}_{l}(\vec{r})d\vec{r}}} =
  \sum_{k,l} {G_{lk}
  [\bfm{S}_{\rm M}]^{-1}_{kl}  }.
 \end{split}
\eeque
 The LDOS is thus given by
\bequ
D^{\Delta}_{\rm M}(\omega) = -\frac{2}{\pi} Im \left\{
  Tr \left[\bfm{G}_{\rm M}(\omega)
  [\bfm{S}_{\rm M}]^{-1}  \right] \right\}. 
\label{EQ0046}
\eequ
Notice that Eq. \ref{EQ0046} is different from Eq. \ref{D1}.
Likewise, for the alternative block-orthogonal metric $\nabla$ one gets
\beque
\begin{split}
\rho^{\nabla}_{\rm M}=-\frac{2}{\pi}\int^{0}_{-\infty} { Im \left\{
   Tr \left[
   \bfm{\tilde{G}}_{\rm M}(\omega) \bfm{S}^{*}_{\rm M}  \right] \right\} d\omega},
 \end{split}
\eeque
from which we may define
\bequ
D^{\nabla}_{\rm M}(\omega) = -\frac{2}{\pi} Im \left\{
  Tr \left[\bfm{\tilde G}_{\rm M}(\omega)
  \bfm{S}_{\rm M}^*  \right] \right\}. 
\label{EQ0047}
\eequ
 
\begin{figure}[t]
    \includegraphics*[width=0.45\textwidth]{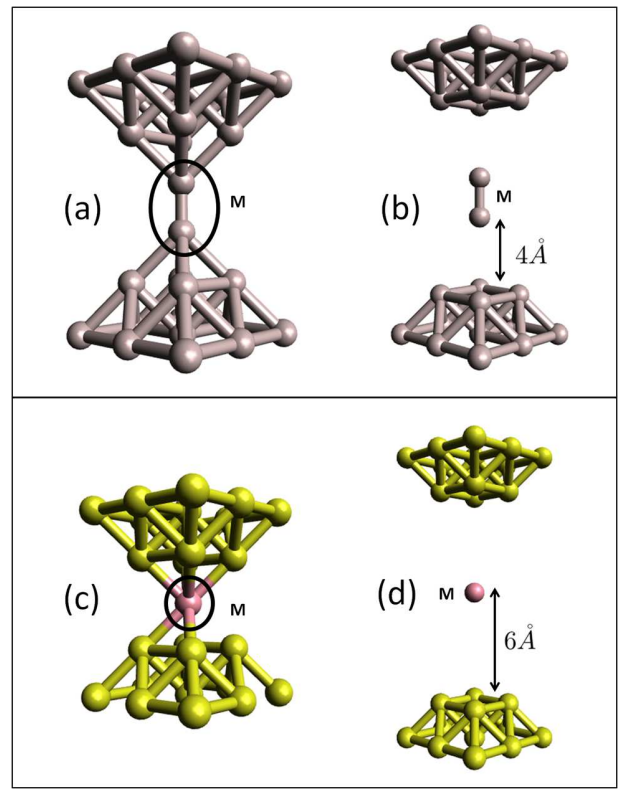} 
  \caption{
    (a-b) View of an aluminum nanocontact for two different connections between the central region M and the electrodes.
    (c-d) The same for a gold nanocontact with a cobalt atom embedded. 
  }
   \label{Fig02}
\end{figure}

\begin{table}[t]
\begin{tabular*}{0.48\textwidth}{ @{\extracolsep{\fill}} 
| c | c | c | c | c | c | c | c | c |} 
   \hline
   & Q & $\rho_{\rm M}$ & $\rho^{\nabla}_{\rm M} $ & $\rho^{\Delta}_{\rm M} $ & $ D_{\rm M} $& 
	$\rho^{*}_{\mathcal{M}} $ & $\rho_{\mathcal{M}} $ & $ D_{\mathcal{M}} $
	\\
  \hline 
		Al(1) & 6 & 6.38  & 2.89 & 11.39 & 16 
	& 6.03 & 34.94 & 45.16 \\
   \hline  
	Al(2) & 26 & 24.03  & 22.52 & 26.87 & 36 
	& 23.52 & 30.44 & 41.86 \\
   \hline
    Al(3) & 26 & 23.20  & 22.76 & 35.10 & 72 
	& 25.83 &  410.49 & 495.00 \\
   \hline  
  Co(1)& 17 & 12.96  & 15.26 & 24.66 & 44 & 
	22.82 & 753.18 & 1043.58 \\
 \hline 	
  Co(2)& 27& 26.61 & 23.95 & 29.26 & 36 & 
	26.04 & 57.64 & 72.34 \\ 
  \hline   
	Co(3)&27& 22.73 & 25.04 & 33.24 & 68 & 
	33.31 & 678.84 & 865.62 \\
   \hline 
    \hline
    Al(4) & 26 & 24.68  & 24.47 & 24.89 & 36 
	& 24.57 & 25.00 & 36.24 \\
   \hline 
		Co(4)&17& 15.94 & 16.26 & 19.02 & 44 & 
	18.05 & 58.12 & 123.36 \\
  \hline 
		Co(5)&17& 16.95 & 16.26 & 17.27 & 44 & 
	17.06 & 17.87 & 50.56 \\
  \hline 
\end{tabular*}
\caption {Partial electronic charge in the central region (M) for four different systems, different basis sets, and the different projections discussed in the text (the metric dimension of these is also shown). For all atoms which are not part of the M region we have considered a minimal basis set ($sp$)\cite{crenbs}. For the Al nanocontacts, cases (1), (2), and (3) correspond to Fig. \ref{Fig02}(a) with the same minimal basis set for the M region, the STO-3G basis set, and the 6-31G$^{**}$ basis 
set, respectively. Case (4) corresponds to Fig. \ref{Fig02}(b) and the STO-3G basis set for the M region. For the Au-Co nanocontacts we consider the LANL2DZ pseudopotentials basis set [(1)], the STO-3G basis set [(2)], and the 6-31G$^{**}$ basis set [(3)] for the Co atom in the geometry shown in Fig. \ref{Fig02}(c). Cases (4) and (5) correspond to Fig. \ref{Fig02}(d) (two different distances, but only (5) shown) and the basis set LANL2DZ. Spin-polarized calculations are performed in all last five cases, but only the sum of spin-up and spin-down charges is shown. }
\label{table01}
\end{table}

To illustrate the different results obtained from the different projections and the different basis sets (direct and block- orthogonal) we have considered two simple models of metallic nanocontacts. The first one is an Al nanocontact with two slightly different atomic structures as shown in Figs. \ref{Fig02}(a) and (b). They differ in how strongly coupled the two central atoms are to the rest of the system. The other model is a Au nanocontact with a Co atom in the middle [Figs. \ref{Fig02}(c) and (d)]. Again, two slightly different geometries with different overlaps between the central atom and the electrodes are considered.
The selected region or subspace for projection, M,  is indicated in both figures. The DFT calculations have been performed  with our ANT.G code \cite{ANT:G} which is based on the non-equilibrium Green's function formalism  and interfaces with GAUSSIAN \citep{g09}. We have chosen a standard exchange-correlation functional in the generalized gradient approximation\citep{becke:pra:88, PhysRevLett.77.3865} for all calculations and various basis sets explained in the caption of Table \ref{table01}. 

Table \ref{table01} shows the results for the charges associated to the different projections along with the dimension of the projected subspace.  As a reference, the charge $Q$ of the isolated M subspaces (when they are infinitely far apart from the electrodes and therefore neutral) is also shown (second column). As can be appreciated, the projected charges lie in a wide range of values. The first result worth noticing is that the values of $D_{\rm M}$ increase to insanely large number as the basis set dimension increases. This is particularly notorious for the Al(3), Co(1), and Co(3) cases. This translates into unrealistic values of $\rho_{\mathcal{M}}$. Interestingly, the values of    $\rho_{\mathcal{M}}^*$ are very reasonable. The other three values for the charge, $\rho_{\rm M}$, $\rho_{\rm M}^\Delta$ and $\rho_{\rm M}^\nabla$, are all within acceptable limits, although still present a large dispersion. In the case of nanocontacts with a single element (Al) these values can be directly  compared to the values of $Q$ since we do not expect a significant charge transfer between the electrodes and the M subspace. It is worth noticing that the Mulliken value for the charge, $\rho_{\rm M}$, generally lies in between the other two values obtained from the  two different block orthogonal metrics $\Delta$ and $\nabla$. The last three rows present results where the M region has been separated from the electrodes, reducing thus the overlap. The differences between all values are consequently reduced as well, including  $\rho_{\mathcal{M}}$. For Co a larger distance than for Al is required to get similar values for all types of projections. 

\section{Reduced Green's functions and effective Hamiltonians}
\label{reduced_green}
\noindent We are now fully equipped to obtain the  effective Hamiltonian associated to the M subspace, i.e, what we have called in the introduction the active region. Our first aim is to obtain an expression similar to the one in Eq. \ref{EQ0032}, but projected onto this region. Then one could read out the Hamiltonian and the associated self-energy from it.  The dual Hermitian projection of the Green's function operator onto this subspace, as defined in Eq. \ref{dualprojection}, is
\beque
\bfm{\hat{P}}_{\rm M}\bfm{\hat{G}} 
\bfm{\hat{P}}_{\rm M}^\dagger = 
 \bfm{\hat{P}}_{\rm M}\left[
    \left(
 \omega - \mu
 \right)
 \bfm{\hat{I}} - \bfm{\hat{H}} 
  - \bfm{\hat{\Sigma}}\right]^{-1}
   \bfm{\hat{P}}^\dagger_{\rm M}
\eeque 
or equivalently
\beque
 \left[\bfm{\hat{P}}_{\rm M}\bfm{\hat{G}} 
\bfm{\hat{P}}_{\rm M}^\dagger \right] ^{-1}=
\left[
\bfm{\hat{P}}_{\rm M}\left[
    \left(
 \omega - \mu
 \right)
 \bfm{\hat{I}} - \bfm{\hat{H}} 
  - \bfm{\hat{\Sigma}}\right]^{-1}
   \bfm{\hat{P}}^\dagger_{\rm M}\right]^{-1}.
\eeque
Notice that 
\beque
 \left[\bfm{\hat{P}}_{\rm M}\bfm{\hat{G}} 
\bfm{\hat{P}}_{\rm M}^\dagger \right] ^{-1} \neq \bfm{\hat{P}}^\dagger_{\rm M}\left[\bfm{\hat{G}} 
 \right] ^{-1}\bfm{\hat{P}}_{\rm M},
\eeque
but we can always write
\bequ
 \left[\bfm{\hat{P}}_{\rm M}\bfm{\hat{G}} 
\bfm{\hat{P}}_{\rm M}^\dagger \right] ^{-1}= 
    \left(
 \omega - \mu
 \right)
 \bfm{\hat{I}}_{\mathcal{M}} - \bfm{\hat{H}}_{\mathcal{M}} 
  - \bfm{\hat{\Sigma}}_{\mathcal{M}},
   \label{pgp}
\eequ
where the first two terms on the right-hand side are
now direct projections
\beque
\bfm{\hat{I}}_{\mathcal{M}} = \bfm{\hat{P}}^\dagger_{\rm M}\bfm{\hat{I}}\bfm{\hat{P}}_{\rm M}
\eeque
\beque
\bfm{\hat{H}}_{\mathcal{M}} = \bfm{\hat{P}}^\dagger_{\rm M}\bfm{\hat{H}}\bfm{\hat{P}}_{\rm M}
\eeque
while the third one is given by
\beque
\bfm{\hat{\Sigma}}_{\mathcal{M}} = \bfm{\hat{P}}^\dagger_{\rm M}\bfm{\hat{\Sigma}}
\bfm{\hat{P}}_{\rm M} + \bfm{\hat{\Sigma}}^{'}_{\mathcal{M}}.
\eeque 
We need a new term added to the projected self-energy operator which is thus defined by 
\bequ 
\bfm{\hat{\Sigma}}^{'}_{\mathcal{M}}= \left[\bfm{\hat{P}}_{\rm M}\bfm{\hat{G}} 
\bfm{\hat{P}}_{\rm M}^\dagger \right] ^{-1} - \bfm{\hat{P}}^\dagger_{\rm M}\left[\bfm{\hat{G}} 
 \right] ^{-1}\bfm{\hat{P}}_{\rm M}.
 \label{sigmaprime}
 \eequ
where both direct and dual projections appear in the definition. Making use now of the definition of inverse, we carry out the projection and inversion in the left-hand side of Eq. \ref{pgp}:
\beque
\begin{split}
\left[\bfm{\hat{P}}_{\rm M}\bfm{\hat{G}} 
\bfm{\hat{P}}_{\rm M}^\dagger\right]^{-1} = 
\left[\sum_{m,n} 
\ketq m \ket \tilde{G}_{mn}  \bra n \braq \right]^{-1} = &
\\
\sum_{m,n} 
\ketq m^{*} \ket [\bfm{\tilde{G}}_{\rm M}]^{-1}_{mn} \bra n^{*} \braq, 
\end{split}
\eeque
so that Eq.  \ref{pgp} becomes 
\beque
 [\bfm{\tilde{G}}_{\rm M}]^{-1}
 =
 \left(
 \omega - \mu
 \right) \bfm{S}_{\rm M}
 - \bfm{H}_{\rm M}
 - \bfm{\Sigma}_{\rm M}
\eeque
when represented in the direct basis.
Finally we can obtain the self-energy matrix:
 \bequ
\bfm{\Sigma}_{\rm M} 
 =
 \left(
 \omega - \mu
 \right) \bfm{S}_{\rm M}
 - \bfm{H}_{\rm M}
 -  [\bfm{\tilde{G}}_{\rm M}]^{-1}.
 \label{goodsigma}
 \eequ
If required, we can also obtain $\bfm{\Sigma'}_{\rm M}$ from Eq. \ref{sigmaprime}. This way of obtaining the self-energy associated to any subspace is often called ``reversal engineering'' \citep{jacob:prb:10b}.
Notice that Eq. \ref{goodsigma} is somewhat expected by looking at Eq. \ref{gtilde}, but  our projector theory gives us a rigorous way of obtaining it.  Following Eq. \ref{P*P} one could also apply the alternative projection $\bfm{\hat{P}}^\dagger_{\rm M}\bfm{\hat{G}} \bfm{\hat{P}}_{\rm M}$ and arrive at the following representation in direct space
\beque
 [\bfm{G}_{\rm M}]^{-1}
 =
 \left(
 \omega - \mu
 \right) \bfm{S}^{-1}_{\rm M}
 - \bfm{\tilde{H}}_{\rm M}
 - \bfm{\tilde{\Sigma}}_{\rm M}
\eeque
from which we obtain another expression for the self-energy matrix:
\bequ
\bfm{\tilde{\Sigma}}_{\rm M}
 =
 \left(
 \omega - \mu
 \right) \bfm{S}^{-1}_{\rm M}
 - \bfm{\tilde{H}}_{\rm M}
 -  [\bfm{G}_{\rm M}]^{-1}.
\label{badsigma}
 \eequ
 
Finally, the self-energy matrix in the case of the two introduced block orthogonal metrics should be defined by
\bequ
\bfm{\Sigma}^{\Delta}_{\rm M}
 =
 \left(
 \omega - \mu
 \right) \bfm{S}_{\rm M}
 - \bfm{H}_{\rm M}
 -  \bfm{S}_{\rm M}[\bfm{G}_{\rm M}]^{-1}\bfm{S}_{\rm M},
\label{sigmadelta} 
\eequ
and
\bequ
\bfm{\Sigma}^{\nabla}_{\rm M} =
 \left(
 \omega - \mu
 \right) \bfm{S}^{-1}_{\rm M}
 - \bfm{\tilde{H}}_{\rm M}
 -  \bfm{S}^{-1}_{\rm M}[\bfm{\tilde{G}}_{\rm M}]^{-1}\bfm{S}^{-1}_{\rm M}.
\label{sigmanabla} 
 \eequ
Notice that the expressions for the self-energy in Eqs. \ref{goodsigma}, \ref{badsigma}, \ref{sigmadelta}, and \ref{sigmanabla}   are  all different. In conclusion, the choice of
projection for a fully non-orthogonal basis set (direct or dual) or the choice of the block-orthogonal basis set ($\Delta$ or $\nabla$) determines both a Hamiltonian and an associated self-energy. These two must go together for the evaluation of, e.g., conductance as next section shows.

\section{The transmission function from a reduced Green's function}
\label{conductance}

The transmission function $T$, which enters Landauer's formalism to compute the conductance, $G=\frac{2e^2}{h} T$, is given by:
\beque
T =
 Tr \left[\bfm{t}^{\dagger}\bfm{t}\right]
\eeque
where $\bfm{t} = \bfm{\Gamma}_{\rm R}^{1/2}\bfm{\tilde G}^{(+)}\bfm{\Gamma}_{\rm L}^{1/2}$ and $\bfm{t}^{\dagger} =
\bfm{\Gamma}_{\rm L}^{1/2}\bfm{\tilde G}^{(-)}\bfm{\Gamma}_{\rm R}^{1/2}$ are transmission matrices defined in terms of retarded (+) and advanced (-) Green's functions and coupling matrices to the right (R) and left (L) electrodes. We are assuming here that there is no direct electronic coupling between electrodes and that the self-energy $\bfm\Sigma $ that appears in the Green's function  (see Eq. \ref{gtilde})  is known and equal to $\bfm{\Sigma}_{\rm R}+\bfm{\Sigma}_{\rm L}$. The coupling matrices are defined as usual: $\bfm{\Gamma}=i\bfm{(\Sigma^{(+)}-\Sigma^{(-)})}$.  The transmission matrices $\bfm{ t^{\dagger}}$ and $\bfm{t}$
contain information about transmission amplitudes between orbitals or basis elements at the entrance and exit of the scattering region.  
By using the cyclic property of the trace, the transmission function can thus be written in its most popular form:
\bequ
\begin{split}
T
=  Tr \left[\bfm{{\Gamma}}_{\rm L}\bfm{\tilde{G}^{(-)}}
\bfm{{\Gamma}}_{\rm R}\bfm{\tilde{G}^{(+)}}\right],
\label{caroli}
\end{split}
\eequ
Notice that the final product of four matrices \textit{is not} simply the result of representing the product of operators
\beque
\bfm{\hat{\Gamma}}_{\rm L}\bfm{\hat{G}^{(-)}}
\bfm{\hat{\Gamma}}_{\rm R}\bfm{\hat{G}^{(+)}},
\eeque
in a non-orthogonal basis set (see Sec. \ref{basics}) .

We are now interested in expressing the transmission function in terms of Green's functions and coupling matrices reduced or projected to a central scattering region. As an example we will compute the conductance of the Al nanocontact studied in Sec. \ref{projected_charge} where the central region M is the same used to analyse the projection of the charge [see Figs. \ref{Fig02}(a) and (b)].  First, to obtain the self-energy representing the left electrode, we can perform a projection onto the rest of the system:
\beque
\bfm{\hat{P}}_{\rm r}\bfm{\hat{G}} 
\bfm{\hat{P}}_{\rm r}^\dagger =  
 \bfm{\hat{P}}_{\rm r}\left[
    \left(
 \omega - \mu
 \right)
 \bfm{\hat{I}} - \bfm{\hat{H}} 
  - \bfm{\hat{\Sigma}}_{\rm L} - \bfm{\hat{\Sigma}}_{\rm R}\right]^{-1}
   \bfm{\hat{P}}^\dagger_{\rm r}.
\eeque 
Following Eq. \ref{goodsigma} we obtain a new left self-energy
 \beque
 \bfm{\Sigma}_{\rm L|r} 
 = 
 \left(
 \omega - \mu
 \right) \bfm{S}_{\rm r}
 - \bfm{H}_{\rm r} - \bfm{\Sigma}_{\rm R|r}
 -  [\bfm{\tilde{G}}_{\rm r}]^{-1},
 \eeque
where $\bfm{\hat{P}}^\dagger_{\rm r}\bfm{\hat{\Sigma}}_{\rm R}
\bfm{\hat{P}}_{\rm r}:=\bfm{\Sigma}_{\rm R|r}$. 
Then, by performing a new projection onto the selected subspace M we obtain the new self-energy associated to the new right electrode:
\beque
\begin{split}
& \bfm{\hat{P}}_{\rm M}\bfm{\hat{G}}_{\rm r} 
\bfm{\hat{P}}_{\rm M}^\dagger =  \\
&
 \bfm{\hat{P}}_{\rm M}\left[
    \left(
 \omega - \mu
 \right)
 \bfm{\hat{I}}_{\rm r} - \bfm{\hat{H}}_{\rm r} - \bfm{\hat{\Sigma}}_{\rm R|r}
  - \bfm{\hat{\Sigma}}_{\rm L|r}\right]^{-1}
   \bfm{\hat{P}}^\dagger_{\rm M},
\end{split}
\eeque
 \beque
\bfm{\Sigma}_{\rm R|M} 
 =
 \left(
 \omega - \mu
 \right) \bfm{S}_{\rm M}
 - \bfm{H}_{\rm M} - \bfm{\Sigma}_{\rm L|M}
 -  [\bfm{\tilde{G}}_{\rm M}]^{-1},
 \eeque 
 where $\bfm{\hat{\Sigma}}_{\rm L|M} = \bfm{\hat{P}}^\dagger_{\rm M}\bfm{\hat{\Sigma}}_{\rm L|r}\bfm{\hat{P}}_{\rm M}$.
 Now the newly obtained $ \bfm{\tilde{G}}_{\rm M}$, $\bfm{\hat{\Sigma}}_{\rm L|M}$, and $\bfm{\hat{\Sigma}}_{\rm R|M}$ enter Eq. \ref{caroli}. This double-projection procedure has been illustrated through the dual Hermitian projection, but the same operations can be carried out with the other projections and with the block-orthogonal basis sets, as explained in previous section.
\begin{figure}[t]
\includegraphics[width=1.0\linewidth]{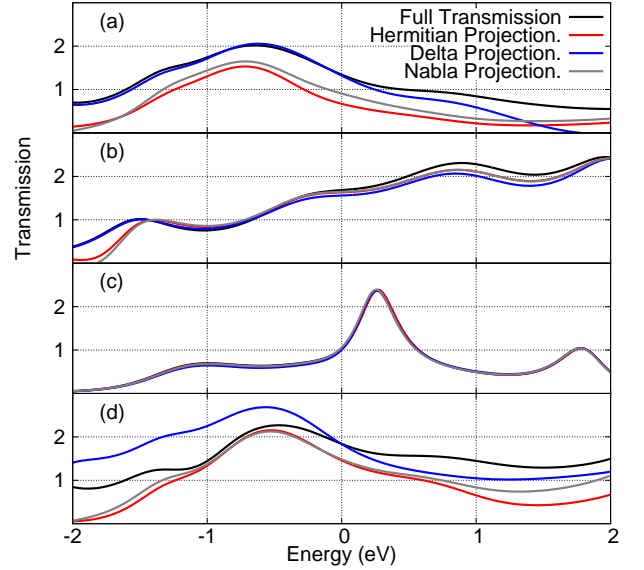} 
\caption {Transmission function for an Al nanocontact as obtained from the full Green's function (black solid line) and from reduced Green's functions with three different projections. Results obtained with the STO-3G$^{**}$ basis set for all atoms are shown in panels (a-c) while results obtained using the 631-G$^{**}$ basis set in the central region are shown in panel (d). The distance between the electrodes increases from (a) [see Fig. \ref{Fig02}(a)] to (c) [see Fig. \ref{Fig02}(b)] while the atomic structure corresponding to panel (d) is the same as for panel (a).} 
\label{FIG03}
\end{figure}

In Figs. \ref{FIG03}(a) and (c) we plot the transmission function obtained for the systems shown in Fig. \ref{Fig02}(a) and (b). We also show the transmission for the same system with an intermediate distance [$d=3$\AA, Fig. \ref{FIG03}(b)] between the central region and the electrodes. The same basis set has been used in the panels (a-c). Panel (d) shows the transmission with a larger basis set for the structure in Fig. \ref{Fig02}(a).
Three different projections have been used in all cases, as described by Eq. \ref{goodsigma}, \ref{sigmadelta}, and \ref{sigmanabla}. The reduced Green's functions and coupling matrices in Eq. \ref{caroli} have been chosen accordingly. When the central region is separated from the electrodes [see Fig. \ref{FIG03}(c)], the direct tunnelling between electrodes is negligible and the reduction operation is expected to be exact. In fact, the three
projections give identical results and equal to the transmission obtained from the full Green's function (black solid line). This corroborates the mathematical procedure of reduction explained in previous section. On the other hand, when direct tunnelling between electrodes is finite, differences are expected between the full transmission and the ones obtained after reduction. This can be clearly seen in Figs. \ref{FIG03}(a), (b), and (d).

\section{Conclusions}
Starting from the very basic definition of metric associated to a given non-orthogonal basis set and the corresponding closure relation, we have defined two types of projection operators.  In doing so, the interplay between direct and dual spaces has been stressed. We have shown how these two projectors can be used to obtain different types of subspaces with different metrics, two with non-integer dimension and one with integer dimension. Physical quantities such as local charge and LDOS for simple systems such as metallic nanocontacts have been evaluated. Large differences can be obtained depending on the projection chosen and degree of overlap present in the original basis set. We have concluded that what we denote as direct Hermitian projection severely overestimates the charge while the dual one gives reasonable values (as compared to a standard Mulliken population analysis which is obtained through non-Hermitian projections). A comparison with results obtained with a block orthogonal basis (which mixes direct and dual basis sets) has also been presented. The results obtained in this mixed basis set also result in projected charges close to Mulliken charges. We have finally shown how to properly carry out different projections on the Green's function, from which effective Hamiltonians can be derived through reversal-engineering. The evaluation of the conductance using these reduced Green's functions and the corresponding coupling matrices certifies the consistency of our reduction or projection procedures.

\section*{Acknowledgments}
This work was supported by MICINN under Grants Nos.
FIS2010-21883 and CONSOLIDER CSD2007-0010, by Generalitat Valenciana under Grant PROMETEO/2012/011. The authors acknowledge D. Jacob, L. A. Zotti and G. G\'omez-Santos for enlightening discussions.
The authors thankfully acknowledge the computer resources, technical expertise, and assistance provided by the Supercomputing and Visualization Center of Madrid (CeSViMa) and support from the Centro de Computaci\'on Cient\' ifica of the Universidad Aut\'onoma de Madrid.

\bibliographystyle{apsrev}


\end{document}